\documentclass[twoside,twocolumn,9pt]{article}
\usepackage{extsizes}
\usepackage[super,sort&compress,comma]{natbib} 
\usepackage[version=3]{mhchem}
\usepackage[left=1.5cm, right=1.5cm, top=1.785cm, bottom=2.0cm]{geometry}
\usepackage{amssymb}
\usepackage{balance}
\usepackage{mathptmx}
\usepackage{sectsty}
\usepackage{graphicx} 
\usepackage{lastpage}
\usepackage[format=plain,justification=justified,singlelinecheck=false,font=stretch=1.125,small,sf,labelfont=bf,labelsep=space]{caption}
\usepackage{float}
\usepackage{fancyhdr}
\usepackage{fnpos}
\usepackage[english]{babel}
\addto{\captionsenglish}{%
  
}
\usepackage{soul}
\usepackage{array}
\usepackage{droidsans}
\usepackage{charter}
\usepackage[T1]{fontenc}
\usepackage[usenames,dvipsnames]{xcolor}
\usepackage{setspace}
\usepackage[compact]{titlesec}
\usepackage{hyperref}
\usepackage{caption}
\usepackage{subcaption}

\usepackage{epstopdf}

\definecolor{cream}{RGB}{222,217,201}

\begin{document}

\pagestyle{fancy}
\thispagestyle{plain}
\fancypagestyle{plain}{

\renewcommand{\headrulewidth}{0pt}
}

\makeFNbottom
\makeatletter
\renewcommand\LARGE{\@setfontsize\LARGE{15pt}{17}}
\renewcommand\Large{\@setfontsize\Large{12pt}{14}}
\renewcommand\large{\@setfontsize\large{10pt}{12}}
\renewcommand\footnotesize{\@setfontsize\footnotesize{7pt}{10}}
\makeatother

\renewcommand{\thefootnote}{\fnsymbol{footnote}}
\renewcommand\footnoterule{\vspace*{1pt}%
\color{cream}\hrule width 3.5in height 0.4pt \color{black}\vspace*{5pt}} 
\setcounter{secnumdepth}{5}

\makeatletter 
\renewcommand\@biblabel[1]{#1}            
\renewcommand\@makefntext[1]%
{\noindent\makebox[0pt][r]{\@thefnmark\,}#1}
\makeatother 
\renewcommand{\figurename}{\small{Fig.}~}
\sectionfont{\sffamily\Large}
\subsectionfont{\normalsize}
\subsubsectionfont{\bf}
\setstretch{1.125} 
\setlength{\skip\footins}{0.8cm}
\setlength{\footnotesep}{0.25cm}
\setlength{\jot}{10pt}
\titlespacing*{\section}{0pt}{4pt}{4pt}
\titlespacing*{\subsection}{0pt}{15pt}{1pt}

\fancyfoot{}
\fancyfoot[RO]{\footnotesize{\sffamily{1--\pageref{LastPage} ~\textbar  \hspace{2pt}\thepage}}}
\fancyfoot[LE]{\footnotesize{\sffamily{\thepage~\textbar\hspace{3.45cm} 1--\pageref{LastPage}}}}
\fancyhead{}
\renewcommand{\headrulewidth}{0pt} 
\renewcommand{\footrulewidth}{0pt}
\setlength{\arrayrulewidth}{1pt}
\setlength{\columnsep}{6.5mm}
\setlength\bibsep{1pt}

\makeatletter 
\newlength{\figrulesep} 
\setlength{\figrulesep}{0.5\textfloatsep} 

\newcommand{\topfigrule}{\vspace*{-1pt}%
\noindent{\color{cream}\rule[-\figrulesep]{\columnwidth}{1.5pt}} }

\newcommand{\botfigrule}{\vspace*{-2pt}%
\noindent{\color{cream}\rule[\figrulesep]{\columnwidth}{1.5pt}} }

\newcommand{\dblfigrule}{\vspace*{-1pt}%
\noindent{\color{cream}\rule[-\figrulesep]{\textwidth}{1.5pt}} }

\newcommand{\gdot}{\dot{\gamma}}

\newcommand{\mh}[1]{\textcolor{blue}{#1}}

\newcommand{\smf}[1]{\textcolor{red}{#1}}

\makeatother

\twocolumn[
  \begin{@twocolumnfalse}
{\includegraphics[height=30pt]{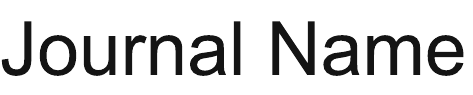}\hfill\raisebox{0pt}[0pt][0pt]{\includegraphics[height=55pt]{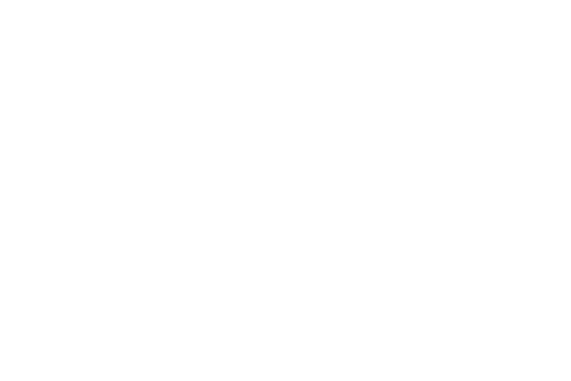}}\\[1ex]
\includegraphics[width=18.5cm]{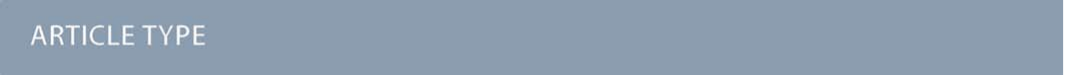}}\par
\vspace{1em}
\sffamily
\begin{tabular}{m{4.5cm} p{13.5cm} }

\includegraphics{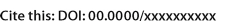} & \noindent\LARGE{\textbf{Viscoelasticity and elastoplasticity in the power law creep and yielding of gels and fibre network materials under stress.$^\dag$}} \\
\vspace{0.3cm} & \vspace{0.3cm} \\

 & \noindent\large{Michael J. Hertaeg,\textit{$^{a}$} Suzanne M. Fielding,\textit{$^{a}$}} \\

\includegraphics{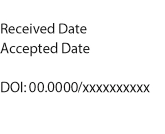} & \noindent\normalsize{We study computationally the creep and yielding of  athermal gels and fibre network materials under a constant imposed shear stress, within a minimal model of interconnected filaments with central forces in $d=2$ spatial dimensions.  Each filament is assumed Hookean initially, then breaks irreversibly above a threshold strain. At early times after the imposition of a small stress, we find purely viscoelastic creep response associated with non-affine deformations within the material, with solid terminal behaviour for a network coordination $Z>2d=4$ and initially floppy response for $Z<4$. For a marginally connected network, $Z=4$, we find sustained power law creep with a strain rate $\gdot\sim t^{-1/2}$ and strain $\gamma \sim t^{1/2}$ as a function of time $t$ after the imposition of the stress. This viscoelastic regime gives way at later times to elastoplastic creep arising from filament breakage, broadening the range of values of $Z$ and time over which power law creep occurs, compared to a network with filament breakage disallowed. This accumulating filament breakage can weaken the network to such an extent that catastrophic material failure then occurs after a long delay, which we characterise. Finally, we consider the implications of viscoelastic versus elastoplastic deformation for the extent to which a material will recover its original shape if the load is removed after some interval of creep.} \\

\end{tabular}

 \end{@twocolumnfalse} \vspace{0.6cm}

  ]

\renewcommand*\rmdefault{bch}\normalfont\upshape
\rmfamily
\section*{}
\vspace{-1cm}

\footnotetext{\textit{$^{a}$~Department of Physics, Durham University, Science Laboratories, South Road, Durham DH1 3LE, UK}}

\footnotetext{\dag~Electronic Supplementary Information (ESI) available}

\section{Introduction}

Materials formed of an interconnected internal network structure are widespread in nature and technology. In a biological context,  networks of filamentous proteins~\cite{storm2005nonlinear}  form the basis of the intracellular actin cytoskeleton~\cite{stricker2010mechanics}, and of extracellular components such as  fibrin~\cite{litvinov2017fibrin} and collagen~\cite{licup2015stress}.  Numerous
soft solids such as colloidal gels~\cite{gado2016colloidal,zaccarelli2007colloidal} and hydrogels~\cite{karobi2016creep} are likewise based on an underlying network structure, as are fibrous materials~\cite{chawla2016fibrous,pan2007fibrous} such as paper~\cite{rosti2010fluctuations} and felt~\cite{stulov2004dynamic}. At a macroscopic level, the mechanical properties of network materials are determined not only by the elastic properties of their constituent fibres or filaments individually, but also by the overall structure of the network as a whole. For example, a prototypical disordered athermal network of interconnected filaments formed of Hookean springs with central forces is known by Maxwell's criterion~\cite{Maxwell} to be rigid in linear response only if the average number of bonds per node $Z>2d$, in $d$ spatial dimensions. In contrast, networks with $Z<2d$ are floppy in linear response, but stiffen when strained~\cite{wyart2008elasticity}.

The manner in which  gels and fibre network materials deform under an imposed load has been widely studied experimentally. In a commonly performed protocol, a shear stress of some amplitude  $\Sigma$ is switched on at time $t=0$ and held constant thereafter. The way a material deforms in response to the applied load is then characterised by its creep curve of shear strain as a function of time, $\gamma(t>0)$. This is often also reported in time-differentiated form to give the evolving shear rate, $\gdot(t>0)$. 

For small imposed stresses, or at early times after the imposition of a larger stress, the shear rate typically decreases over time as a power law, $\gdot \sim t^n$ with $n<0$\cite{lidon2016creep, grenard2014timescales}. For exponent values $-1<n<0$, the strain increases sublinearly as $\gamma \sim t^{n+1}$, giving power law creep. The special case $n=-1$ corresponds to logarithmic creep, $\gamma\sim \log t$.  In this early creep regime, the network filaments typically undergo purely reversible displacements, with the material fully recovering its original shape if the stress is later removed \cite{aime2018power}. These displacements can be partially affine (uniformly following the global deformation), and partially non-affine (with a heterogeneous internal strain field). Collectively, they determine the material's linear viscoelastic response, which can be measured either under imposed stress (our focus here), or imposed strain. Indeed, power law viscoelasticity arises widely in soft materials (network-based or otherwise), including foodstuffs~\cite{ng2008power,mackley1994rheological,leocmach2014creep,faber2017describing}, personal care products~\cite{mackley1994rheological}, paints~\cite{mackley1994rheological}, slurries~\cite{mackley1994rheological}, microgels~\cite{ketz1988rheology}, foams~\cite{khan1988foam}, dense emulsions~\cite{mason1995elasticity}, dense colloids~\cite{mason1995linear}, surfactant onion phases~\cite{panizza1996viscoelasticity}, clays~\cite{hoffmann1993aggregating, rich2011size}, hydrogels~\cite{hung2015fractal}, polymer melts~\cite{mackley1994rheological}, biopolymer networks~\cite{broedersz2010cross}, and biological cells~\cite{kollmannsberger2011linear} and tissues~\cite{davis2006constitutive,khalilgharibi2019stress,hoffman2006consensus}.

At longer times and/or larger stresses, the creep of a network under an imposed load ceases to be purely viscoelastic. Instead, irreversible plastic deformation slowly accumulates within the material\cite{kamani2021unification}. The resulting  macroscopic shape change is then at least partially unrecoverable: a sample will remain permanently deformed to some degree even if the load is later removed. 

Logarithmic and/or power law creep has been observed experimentally in numerous disordered soft materials, including  granular matter~\cite{knight1995density}, colloidal glasses~\cite{siebenburger2012creep}, colloidal gels~~\cite{sprakel2011stress,ballesta2016creep}, surfactant columnar phases~\cite{bauer2006collective}, carbopol gel~\cite{divoux2011stress} and crumpled sheets~\cite{matan2002crumpling,lahini2023crackling}, as well as in harder crystalline or polycrystalline materials such as ice~\cite{glen1955creep,duval1978anelastic}  and  metals~\cite{mclean1966physics,kassner2015fundamentals}. It has also been studied in particle simulations~\cite{chaudhuri2013onset,landrum2016delayed} and elastoplastic models~\cite{popovic2022scaling,liu2009study,korchinski2025thermal}.

The internal plasticity associated with creep can progressively weaken a material to such an extent that the slow creep seen initially after loading is later interrupted by a dramatic yielding process, signified macroscopically by a sudden large increase in the measured strain and strain rate. Catastrophic material failure  after a lengthy induction time  under a constant load has been reported in a myriad of disordered soft solids such as transient 
gels~\cite{poon1999delayed,skrzeszewska2010fracture}, polymeric gels~\cite{bonn1998delayed}, colloidal gels~\cite{gopalakrishnan2007delayed,gibaud2010heterogeneous,grenard2014timescales,sprakel2011stress,cho2022yield,aime2018microscopic,moghimi2021yielding,landrum2016delayed} and hydrogels~\cite{karobi2016creep}; in biological materials such as  protein gels~\cite{leocmach2014creep}, collagen networks~\cite{gobeaux2010power} and biopolymer gels~\cite{pommella2020role}; and in fibrous materials such as  paper~\cite{rosti2010fluctuations} and fibre composites~\cite{nechad2005creep}.    Such behaviour  has important implications for the integrity of materials in applications such as construction, as well as for geophysical phenomena such as mudslides, avalanches and earthquakes.

In this work, we study by numerical simulation the creep and yielding of a minimal model of an athermal disordered network material subject to an imposed shear stress in $d=2$ dimensions. The model and step stress protocol are described in Secs.~\ref{sec:model} and~\ref{sec:protocol} respectively. Sec.~\ref{sec:numerics} introduces a numerical method designed to maintain the shear stress at a constant value over time in our simulations. (Soft materials are more commonly simulated under an imposed shear strain, with the time-evolving stress measured in response.) Sec.~\ref{sec:parameters} summarises units and parameter values and our results  are set out in Sec.~\ref{sec:results}

The contributions of this work are threefold. First, we examine the extent to which the minimal  model studied here can capture the power law creep that is observed so widely in network materials. To this end, we start by characterising creep at early times in the purely viscoelastic regime, with no plastic bond failure events. Consistent with Maxwell's criterion, we observe solid response for $Z>2d=4$ and initially floppy behaviour for $Z<4$, but without sustained power law creep in either case. Only for the particular case of a marginally connected network, $Z=4$, do we find sustained power law viscoelastic creep, with a strain rate $\gdot\sim t^{-0.5}$  and strain $\gamma\sim t^{0.5}$ as a function of time $t$ after the imposition of the stress. We then introduce plastic bond failure events and study creep in the elastoplastic regime. Here,  a gradual process of accumulating bond breakages  widens somewhat the range of values of (initial) network connectivity $Z$ and time $t$ over which power law creep arises. We return in the concluding Sec.~\ref{sec:conclusions} to consider other possible mechanisms for power law creep, not contained in the model studied here. 

Our second contribution will be to characterise the delayed catastrophic material failure that arises at later times or for larger imposed stresses.  Here we report the network's yield stress as a function of the individual local bond breakage strain $b$  and the network's (initial) connectivity $Z$: {\it i.e.}, of the two parameters that respectively determine the individual and collective behaviour of the network's constituent filaments. We also report as a function of $b$ and $Z$ the network's threshold strain, defined as the strain attained at long times at the largest imposed value of imposed stress for which yielding does not occur.  We further characterise the time taken for failure to occur at larger imposed stresses, with a particular focus on the dramatic increase in this timescale as the yield stress is approached from above. Finally, third, we consider the implications of internal nonaffine deformations and plastic bond breakage for the degree to which the strain that arises during creep is recoverable, {\it i.e.}, for the degree to which a sample will return to its original shape if it is unloaded after some interval of creep.

\section{Fibre network model}
\label{sec:model}

To prepare an athermal disordered fibre network in $d=2$ spatial dimensions, we first randomly initialise an ensemble of $N$ athermal soft disks at area fraction $\phi=1.0$ in a square box of size $L\times L$, with periodic boundary conditions. The ensemble comprises equal numbers of small and large disks with  bidisperse  radius ratio $1:1.4$. Any pair of contacting disks  has an elastic Hookean repulsive force. Each disk moves under the net elastic force arising from all its contacts with neighbouring disks, against a dissipative  background characterised by a drag coefficient. With these dynamics, we evolve the packing to steady state. Once a steady state is attained, we take each pair of overlapping disks and connect their centres with a spring, with equilibrium length $l_0$ equal to its initial length $l$. We then remove the disks to leave a random network of springs.  

The network is finally pruned via an iterative process to a desired average connectivity $Z$, defined as the number of springs meeting each node, on average across the network, such that a network of $N$ nodes has $NZ/2$ springs. At each step in the iteration, the network nodes are ranked by the number of connecting springs meeting each node. From the population of nodes with the highest number of connections, we select a node at random and remove  one of its connecting springs at random. We then remove  all rattlers and danglers, defined as nodes with zero and one connecting springs respectively, and their associated springs. This process is repeated iteratively until the network's average connectivity $Z$ attains its desired value. 

Each network spring is assumed Hookean with spring constant $K$ up to a failure length $l$ and corresponding failure strain $b=(l-l_0)/l_0$, beyond which it immediately breaks permanently.

\section{Rheological protocol}
\label{sec:protocol}

After preparing the network in the manner just described, we switch on at time $t=0$ a shear stress of magnitude $\Sigma$ and hold it  constant thereafter, using the numerical algorithm described in Sec.~\ref{sec:numerics}. Our aim is to study the way in which the network creeps and,  for large enough loads,  yields and fails in response to this applied load. We monitor in particular the accumulating macroscopic shear strain across the sample as a whole, as characterised by the creep curve $\gamma(t)$,  and the associated shear rate $\gdot(t)=d\gamma/dt$.

\section{Numerical method}
\label{sec:numerics}

We perform $2d$ shear simulations  using Lees-Edwards periodic boundary conditions, with velocity direction $\hat{x}$ and velocity gradient direction $\hat{y}$,  ignoring the   vorticity direction $\hat{z}$. Driven by  the applied shear and by the net Hookean force  from all  springs connected to it, $\Vec{F}_{\rm elastic}$, each node moves against a dissipative background with drag coefficient $\zeta$.  Accordingly, the position of the $i$th node evolves according to:
\begin{equation}
\label{eqn:dynamics}
  \frac{d\Vec{r}_i}{dt}=\gdot(t) y_i\hat{x}+\frac{1}{\zeta}\Vec{F}_{i,\rm elastic}.
\end{equation}

Simulations of sheared soft materials are most commonly performed at an imposed shear rate $\gdot$. In performing simulations instead at imposed shear stress, we must use an algorithm that appropriately adjusts $\gdot=\gdot(t)$ in order to keep the shear stress  constant. To achieve this, we used two different algorithms, which we call I and II, tested against each other for consistency.

In each algorithm, we introduce for numerical convenience a fictitious Newtonian solvent of viscosity $\eta$,  such that the total shear stress $\Sigma=\sigma+\eta\gdot$.  Our aim is then to perform simulations that effectively attain the limit $\eta\to 0$, with essentially no dependence on $\eta$. Here $\sigma$  is the elastic stress that arises from the network,  calculated at each timestep using the Kirkwood formula 
\begin{equation}
\sigma_{\alpha\beta}=\frac{1}{2A}\sum_{i=1}^N f_{i,\alpha} r_{i,\beta}.
\end{equation}
The sum is performed over all networks nodes, with $f_{i,\alpha}$ the $\alpha$ component of force  and $r_{i,\beta}$ the $\beta$ component of position of node $i$. The sum is normalised by the system area $A$.

\begin{figure}[!t]
\includegraphics[width=\columnwidth]{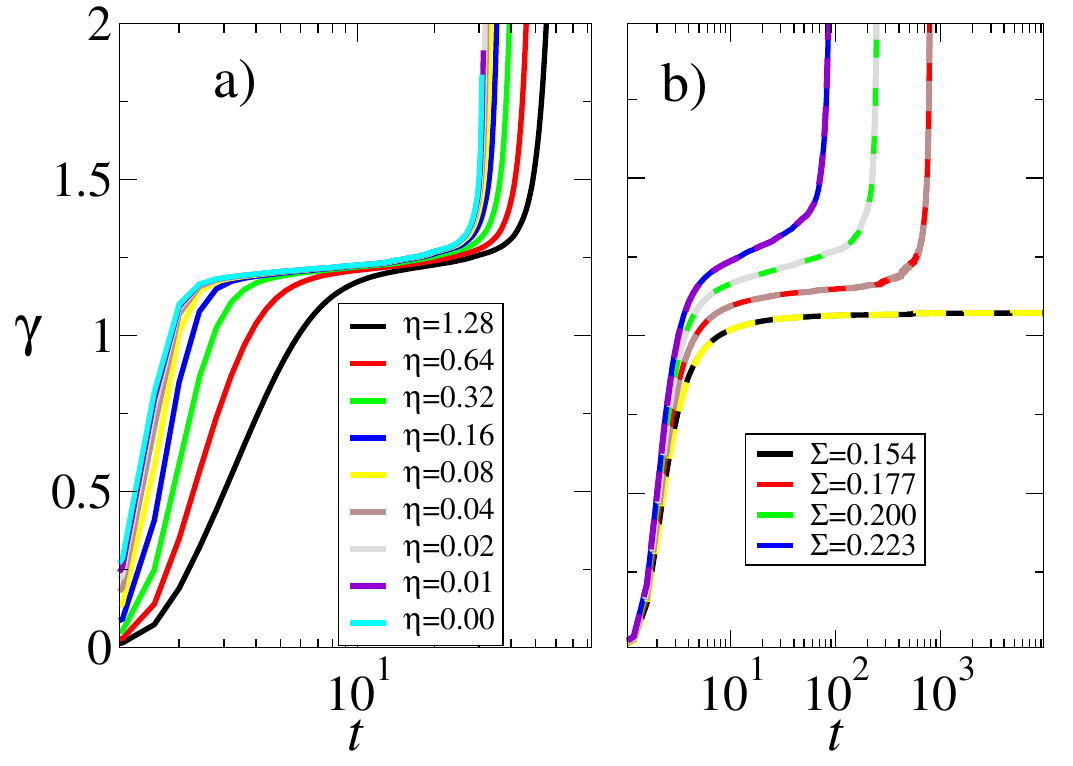}
\caption{Validation of our numerical method for simulating a random network at constant imposed stress. a) Strain versus time using algorithm I for viscosity $\eta=0.00, 0.01, 0.02, 0.04, 0.08, 0.16, 0.32, 0.64, 1.28$ in curves left to right at an imposed stress $\Sigma = 0.18095$. b) Strain versus time computed using algorithm II (solid lines) and algorithm I (dashed lines) for viscosity $\eta = 0.16$ and stress $\Sigma = 0.154, 0.177, 0.200, 0.223$ in curves upward. Filament breakage threshold $b = 1$ and initial network connectivity $Z=3.5$ in both panels.}
  \label{fig:numerics}
\end{figure}

Algorithm I is defined as follows. At each timestep we perform  a trial strain step of size $\gamma_{\rm trail}=\gdot_{\rm trial}\Delta t$ and calculate the resulting stress $\Sigma_{\rm trial}$. This in general differs from the desired imposed value $\Sigma$. We then iterate the value of $\gdot_{\rm trial}$ using the Newton-Raphson method until the desired value  $\Sigma_{\rm trial}=\Sigma$ is achieved, to within an accuracy of 10$^{-10}$. Performing this iteration at every timestep ensures that the stress is held constant for all times $t>0$. Fig.~\ref{fig:numerics}(a) shows the results of simulations performed using this algorithm for a range of viscosity values $\eta$. As can be seen, good convergence to the limit $\eta\to 0$ is attained for  values $\eta\lesssim 0.16$. Indeed,  algorithm I can in fact be implemented even for viscosity  $\eta=0$. However, we also report results for $\eta\neq 0$ in order to test algorithm A directly against algorithm B.

Algorithm II calculates at each timestep the value of the network stress $\sigma$ and then simply imposes a strain rate $\gdot=(\Sigma-\sigma)/\eta$ so that the total stress $\Sigma$ has the desired imposed value. In contrast to algorithm I, algorithm II can be implemented only for non-zero viscosity $\eta$. Fig.~\ref{fig:numerics}(b) shows excellent agreement between algorithms I and II. 

Having shown that algorithms I and II agree, we shall throughout the paper use algorithm II, which is quicker and simpler to implement than algorithm I. Having shown that the limit $\eta\to 0$ is attained to excellent approximation for values $\eta \le 0.16$, we shall use  $\eta = 0.16$ throughout.

\section{Units and parameter values}
\label{sec:parameters}

\begin{table}[!b]
\small
  \caption{Parameters of the network model, rheological protocol and numerical algorithm. Dimensions are expressed in terms of modulus (G), time (T) and length (L).}
  \label{tbl:example1}
  \begin{tabular*}{0.48\textwidth}{@{\extracolsep{\fill}}llll}
    \hline
    Quantity &Symbol    & Dimensions & Value\\
    \hline
    Number of nodes & N & 1 & 16000\\
    Drag coefficient & $\zeta$ & GT/L & 1 (time unit)\\
    Domain length & L & L & $\sqrt{N}$ (length unit)\\
    Spring stiffness & K & G & 1 (modulus unit)\\
    Viscosity & $\eta$ & GT & 0.16\\
    Average connectivity & Z & 1 & Varied\\
    Filament breakage strain & b & 1 & Varied\\
    Applied stress & $\Sigma$ & G & Varied\\
    Timestep & $\Delta t$ & T & 0.01\\
    \hline
  \end{tabular*}
\end{table}

The network model, rheological protocol and numerical method just described have several parameters. These are summarised in Table.~\ref{tbl:example1}. We choose units of mass, time and length  by setting the spring constant $K=1$, the drag coefficient $\zeta=1$ and the box size $L=\sqrt{N}$, such that the typical spring length is $O(1)$. We then set the number of network nodes as large as possible, $N=16,000$,  while still maintaining a feasible simulation time. We use a numerical timestep $\Delta t=0.01$, which gives results converged to the limit $\Delta t\to 0$. The Newtonian viscosity $\eta=0.16$, as just discussed in Sec.~\ref{sec:numerics}. The important physical parameters left to explore  are then  the network connectivity $Z$, the spring breakage strain $b$, and the applied stress $\Sigma$.

\section{Results}
\label{sec:results}

In presenting our results, we start in Sec.~\ref{sec:viscoelastic} with  networks in which filament breakage is disallowed, $b\to\infty$, such that no plastic deformation can arise. Here the creep response is purely viscoelastic. It arises from an initial largely affine shearing of the material shortly after the stress is imposed, followed by a slower non-affine adjustment of the internal network structure at later times. The strain finally saturates to a constant in steady state. In Sec.~\ref{sec:elastoplastic}, we allow plastic deformation by considering finite values of the filament breakage threshold $b$. At early times and small strains, before any filament breakage arises in practice, the initial creep response is again purely viscoelastic. At longer times and larger strains, filaments break and plastic deformation arises. This results in an additional elastoplastic contribution to the creep response, beyond the purely viscoelastic behaviour seen for $b\to \infty$. For large enough imposed stresses, this elastoviscoplastic creep gives way finally to a divergent strain, signifying material failure. 

\subsection{Viscoelastic creep without filament breakage}
\label{sec:viscoelastic}

\begin{figure}[!t]
\includegraphics[width=\columnwidth]{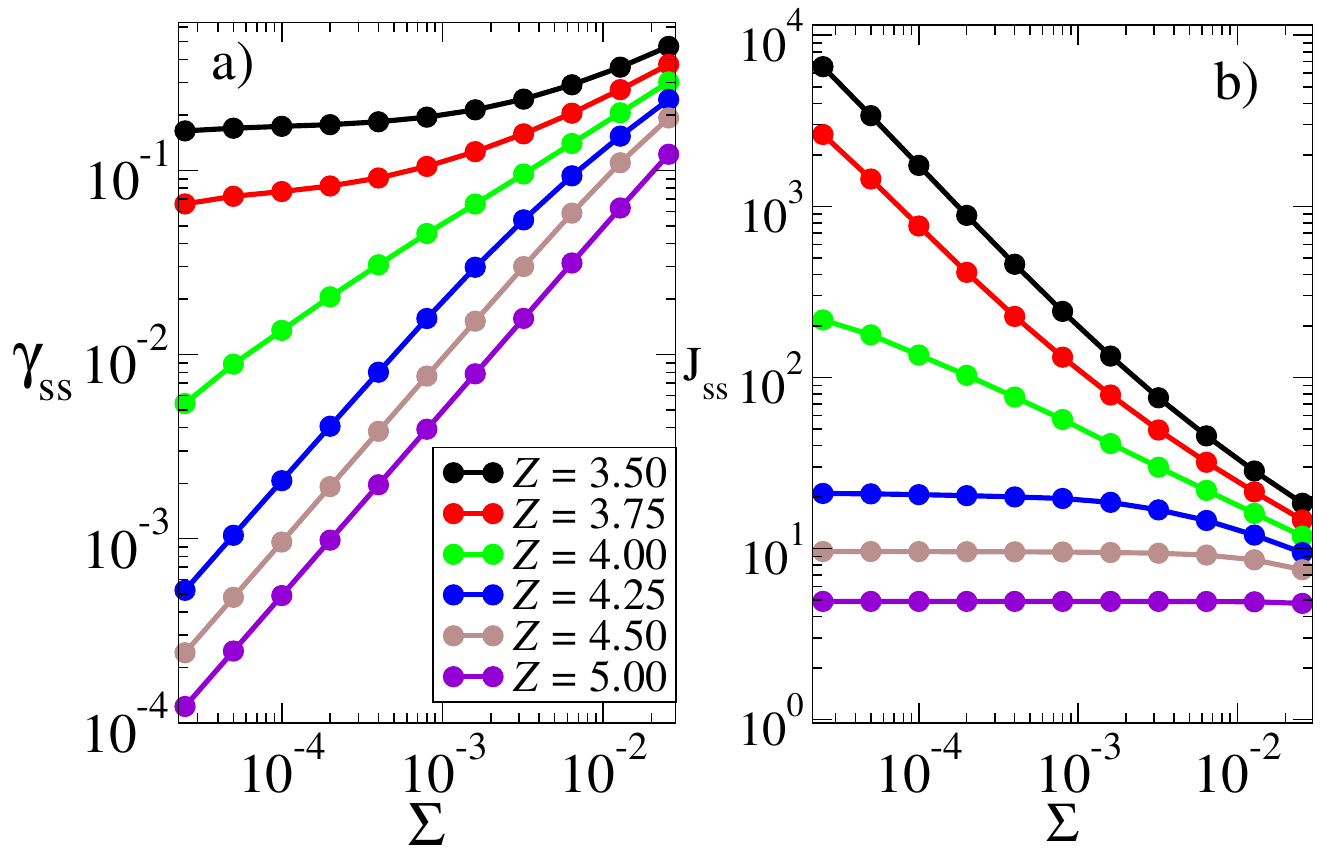}
\caption{(a) Steady state strain attained at long times $t\to\infty$ as a function of imposed stress $\Sigma$, without filament breakage, $b\to\infty$. (b) Corresponding steady state compliance $J_{\rm ss}=\gamma_{\rm ss}/\Sigma$ for the same values of $Z$ as in the left panel.}
\label{fig:steady}
\end{figure}

In this section, we disallow plastic deformation by setting the filament breakage threshold $b\to\infty$. The only physical parameters to explore are then the network connectivity $Z$ and imposed stress $\Sigma$. For any pair of values of these, we prepare a network of connectivity $Z$ using the method described in Sec.~\ref{sec:model} then impose at time $t=0$ a stress $\Sigma$, which is held constant thereafter. We measure the network's strain response to this imposed load, as characterised by its creep curve $\gamma(t)$, and the corresponding strain rate $\gdot(t)$. At long times $t\to\infty$, the strain  rate $\gdot\to 0$ and the strain attains a steady state value $\gamma_{\rm ss}$. This steady state strain is plotted as a function of stress $\Sigma$ in Fig.~\ref{fig:steady}, for several values of connectivity $Z$.

Evident in Fig.~\ref{fig:steady} are two qualitatively different regimes, according to whether the network connectivity $Z>4$ or $Z<4$. For $Z>4$, the steady state strain is proportional to the imposed stress, $\gamma_{\rm ss}\propto \Sigma$, at least in the limit of small stress $\Sigma\to 0$. The material's steady state response  is accordingly that of a Hookean solid. This is consistent with Maxwell's criterion for networks of nodes connected by central forces, which predicts rigidity for a connectivity $Z>2d=4$, in the $d=2$ dimensional simulations performed here.

In contrast, for $Z<4$ the steady state strain $\gamma_{\rm ss}$ attains a non-zero constant 
even in the limit $\Sigma\to 0$: the material is significantly  deformed even by a vanishingly small imposed stress. This nonlinear response marks a significant departure from Hooke's law and is consistent with Maxwell's criterion:  a network of components with purely central forces is floppy for $Z<2d$. Indeed, the results in Fig.~\ref{fig:steady} accord with those of earlier works that reported the steady state stress as a function of increasing  applied strain:  for $Z<4$ the material showed floppy response with zero stress, until a shear-induced stiffening transition at a critical strain $\gamma_{\rm c}(Z)$. Essentially the same behaviour is seen in our Fig.~\ref{fig:steady}, but here with axes inverted, showing strain as a function of imposed stress.

\begin{figure}[!tb]
\centering
\includegraphics[width=1.0\columnwidth]{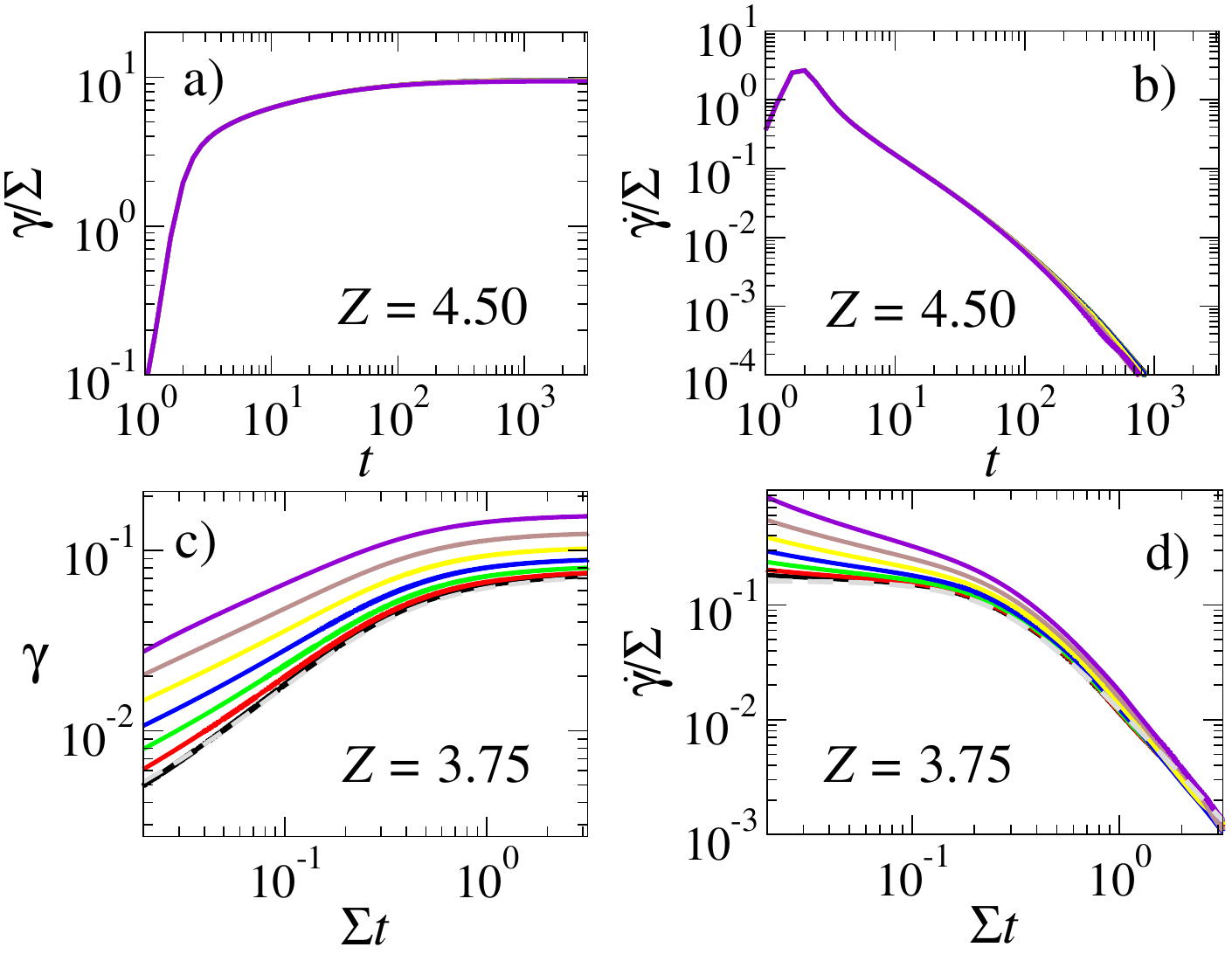}
  \caption{Creep without filament breakage, $b\to\infty$. Strain (left panels a+c) and strain rate (right panels b+d) against time for several imposed stress values $\Sigma = {0.00005, 0.0001, 0.0002, 0.0004, 0.0008, 0.0016, 0.0032}$ in curves from black to violet. Network connectivity $Z=4.50$ (top panels a+b, curves for different stresses indistinguishable) and $Z=3.75$ (bottom panels c+d). Axes have been scaled in each panel to achieve data collapse to a master scaling curve in the limit $\Sigma\to 0$, as discussed in the main text. Accordingly, the raw strain $\gamma(t)$ is shown in panel (c) and the normalised strain, ie, the compliance $J(t)=\gamma(t)/\Sigma$ in panel (a).}
  \label{fig:viscoelastic_sigma}
\end{figure}

So far, then, we have shown that our simulations reproduce at long times $t\to\infty$ the steady state relationship between stress and strain found in previous works. Our primary interest here, however, is in the {\em dynamical} creep response $\gamma(t)$ as a function of time $t$, before that steady state is attained as $t\to\infty$. This is shown in Fig.~\ref{fig:viscoelastic_sigma}, for a representative network connectivity in the stiff regime $Z>4$ in panels (a+b), and a representative connectivity in the floppy regime $Z<4$ in panels (c+d). The normalised strain response $J(t)=\gamma(t)/\Sigma$ gives the compliance.  In the long-time limit, $J(t)$ approaches the steady-state compliance $J_{\rm ss}$ shown in Fig.~\ref{fig:steady}(b).

Before describing these results, we pause to remind ourselves of the key physical quantities in our simulations, and their dimension in terms of modulus $G$, length $L$ and time $t$. We assume no dependence on system size $N$,  solvent viscosity $\eta$, or numerical timestep $\Delta t$, which is valid for large $N$, small $\eta$ and small $\Delta t$. With filament breakage disallowed, $b\to\infty$, the remaining physical quantities are then the spring stiffness $K$ of dimension $G$, the drag coefficient $\zeta$ of dimension $GT/L$, the typical bond length $\Delta=L/\sqrt{N}$ of dimension $L$, the imposed stress $\Sigma$ of dimension $G$, the time $t$ of dimension $T$, the dimensionless strain $\gamma$, and the strain rate $\gdot$ of dimension $T^{-1}$. From these, we can construct the two dimensionless stresses $\Sigma/K=\Sigma$  and $\Sigma t/\zeta \Delta=\Sigma t$,  two dimensionless times $tK/\zeta\Delta=t$ and $t\Sigma/\zeta\Delta=t\Sigma$, and two dimensionless shear rates $\gdot\zeta\Delta/K=\gdot$ and $\gdot\zeta\Delta/\Sigma=\gdot/\Sigma$.

With this in mind, consider now a network with connectivity $Z>4$. Such a network is stiff and will accordingly bear an applied stress, giving a scale of stress $\Sigma \sim K$. In linear response at small strains, the strain will be proportional to stress. Furthermore, the strain must develop on the timescale set by the interaction of dissipation with network stiffness, $tK/\zeta\Delta = t$. Accordingly, we expect the creep curve $\gamma(t)$, measured for several small imposed stresses, to collapse in the limit $\Sigma\to 0$ to a master scaling function when plotted in the format $\gamma(t)/\Sigma$ vs $t$. This is indeed seen in Fig.~\ref{fig:viscoelastic_sigma}(a). The strain rate $\gdot\sim t^{-2}$ at long times $t\to\infty$, as the strain saturates to its steady state constant.

\begin{figure}[!tb]
\includegraphics[width=1.0\columnwidth]{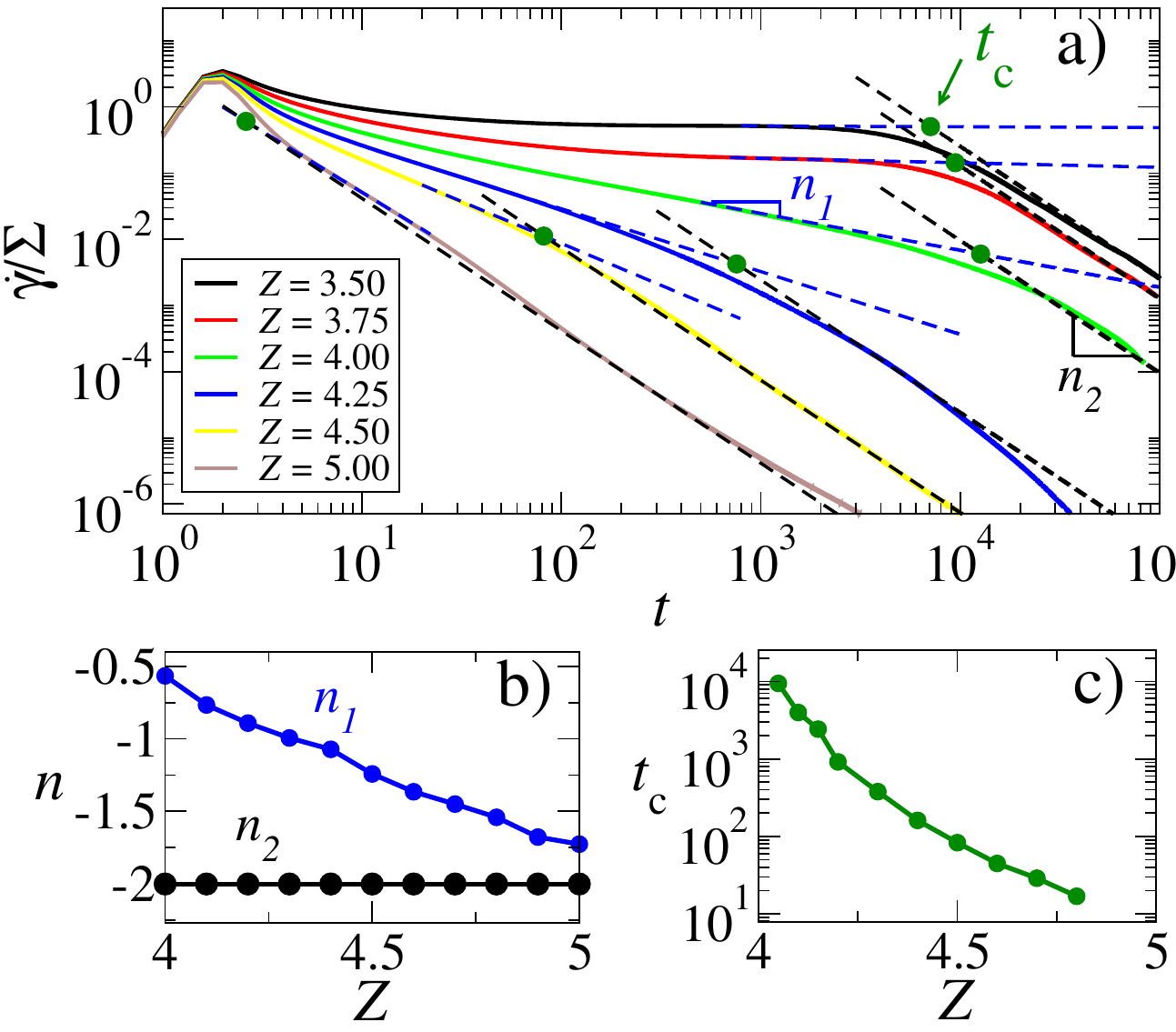}
\caption{Creep without filament breakage, $b\to\infty$. a) Strain rate versus time for an imposed stress $\Sigma = 3.162$x$10^{-5}$ and several values of network connectivity $Z$. Blue and black dashed lines show power law fits to early and late power law regimes respectively. Green dots show crossover points between these two regimes. b) Exponent of early and late power law regimes versus $Z$ c)  Crossover time between early and late regimes versus  $Z$.}
  \label{fig:power}
\end{figure}

In contrast, a network with connectivity $Z<4$ is floppy at small strains and cannot bear an applied load,  until a stiffening transition occurs at finite strain. The stress must therefore be dissipative in nature, scaling as $\Sigma\sim \zeta \Delta/t=1/t$. Likewise, the timescale on which the strain evolves cannot be informed by the network stiffness, and must instead be set by the interaction of dissipation with the imposed load, scaling as $t \Sigma/\zeta\Delta=t\Sigma$. Accordingly, the shear rate $\gdot\sim \Sigma/\zeta\Delta=\Sigma$ and we expect the differentiated creep curves $\gdot(t)$, measured for several small imposed stress values, to converge in the limit $\Sigma\to 0$ to a master scaling function when expressed in the format $\gdot(t)/\Sigma$ versus $t\Sigma$. This is indeed seen in Fig.~\ref{fig:viscoelastic_sigma}(d). Numerical fitting reveals a scaling function of the form
\begin{equation}
\label{eqn:form}
\frac{\gdot(t)}{\Sigma}=\frac{\alpha}{1+\beta(t\Sigma)^2},
\end{equation}
with $\alpha=\alpha(Z)$ and $\beta=\beta(Z)$. This scaling function is shown by the dashed line in d), again with a decay $\gdot\sim t^{-2}$ as $t\to\infty$. Integrating Eqn.~\ref{eqn:form} shows that the corresponding strain attains the master scaling function
\begin{equation}
    \gamma(t)=\frac{\alpha}{\sqrt{\beta}}\arctan(\sqrt{\beta}\Sigma t).
\end{equation}
This indeed agrees with our numerical data for the creep curve in the limit $\Sigma \to 0$, as shown by the dashed line in Fig.~\ref{fig:power}(c). For a floppy network, therefore, the strain depends on the  stress only via the dynamical timescale  $t\Sigma/\zeta\Delta = t\Sigma$, rather than in having a magnitude proportional to it.

So far, we have explored the creep response for a representative network connectivity in the stiff regime,  $Z>4$, and a representative connectivity in the floppy regime, $Z<4$. In Fig.~\ref{fig:power}, we report time-differentiated creep curves $\gdot(t)$ for a range of connectivities, for a fixed small stress. We do so in the representation $\gdot/\Sigma$ vs $t$, suited to the scaling collapse demonstrated in Fig.~\ref{fig:viscoelastic_sigma}(b) for $Z>4$. For $Z<4$, the plateau seen at early times in Fig.~\ref{fig:power} lasts for progressively longer times in simulations repeated for progressively small stresses. 

For network connectivities just above the marginal value $Z=4$, we find two successive power law regimes: $\gdot\sim t^{n_1}$ at early times and $\gdot\sim t^{n_2}$ at late times, separated by a crossover time $t_{\rm c}$. The values of $n_1, n_2$ and  $t_{\rm c}$ are plotted as a function of connectivity $Z>4$ in Figs.~\ref{fig:power}b) and c) respectively. As can be seen, the terminal exponent $n_2=-2.0$ for all $Z$. The early time exponent $n_1\to -0.5$ as $Z\to 4^+$, and with $|n_1|<1$ for $4.0<Z \lessapprox 4.4$. In this regime, the strain grows sublinearly as $\gamma\sim t^{1+n_1}$ with creep exponent $0<1+n_1<1$. The duration $t_{\rm c}$ of this power law creep regime increases dramatically as the connectivity reduces towards the marginal value, $Z\to 4^+$. Indeed,   in a network of infinite size subject to a vanishingly small strain, we expect $t_{\rm c}$ to diverge as $Z\to 4^+$, such that $\gamma/\Sigma$ would grow without bound as $t^{1/2}$ even as $t\to\infty$. In our simulations with finite $N$ and $\Sigma$, $t_{\rm c}(Z=4)=10^4$ is large but finite.  Power law decay of the stress relaxation modulus was observed in a simple network model at marginality in Ref.~\cite{milkus2017atomic}.

In this section, we have explored the viscoelastic creep and eventual steady state strain under an imposed stress in a network in which filament breakage is disallowed, $b\to\infty$. We have shown  in particular that a network of unbreakable springs of connectivity $Z\gtrapprox 4$, just above the marginal value $Z=4$, will show a prolonged regime of power law creep in which the strain increases as $t^{1+n_1}$ with creep exponent $0<1+n_1<1$, up to a crossover time $\tau_{\rm c}$. After that time, the strain rate decreases as $t^{-2}$ and the strain quickly saturates to a constant. The regime of creep explored in this section arises not from elastoplasticity, but from internal non-affine viscoelastic deformations within the network.

\subsection{Elastoplastic creep and yielding via filament breakage}
\label{sec:elastoplastic}

\begin{figure}[!t]
  \includegraphics[width=1.0\columnwidth]{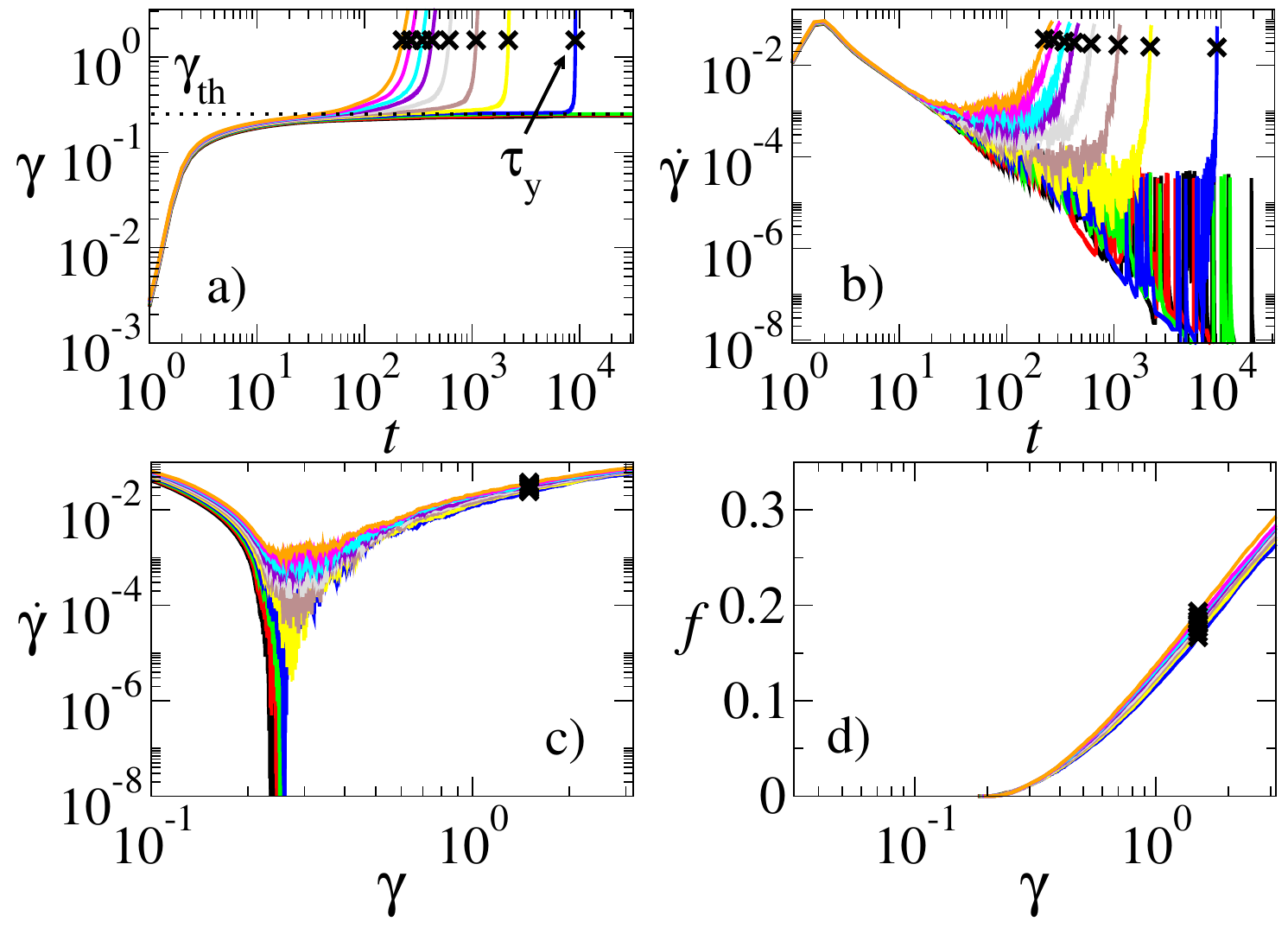}
  \caption{Creep and yielding with filament breakage. a) Strain versus time  for imposed stress values 
$\Sigma=0.0300,0.0305,0.0310,\cdots 0.0350$ 
  in curves right to left. Dashed line indicates the threshold strain $\gamma_{\rm m}$, defined as the strain attained at long times for the largest stress value, $\Sigma=\Sigma_{\rm y}^{-}$, for which yielding does not occur. Crossed indicate the yielding time $\tau_{\rm y}$ for all stress values for which yielding does occur, $\Sigma>\Sigma_{\rm y}$}. b) Differentiated creep curves of strain rate versus time. c) Parametric plot of strain rate versus strain. d) Fraction of filaments broken as a function of strain. Filament breakage threshold  $b = 0.1585$ and initial network connectivity $Z = 4.5$ and  in all panels.
  \label{fig:yielding}
\end{figure}

We now present our results for networks in which filament breakage can occur, with a finite breakage threshold $b$. In this case, the networks's connectivity $Z$ will in general decrease over time after a load is applied. Accordingly, any value of $Z$ quoted in this section refers to the network's initial connectivity in its freshly prepared state prior to loading, unless otherwise stated. 

Fig.~\ref{fig:yielding}(a) shows the strain response of a network of (initial) connectivity  $Z=4.5$ and fixed breakage threshold, subject to several different values of the imposed stress $\Sigma$. At early times and small strains, before any breakage occurs in practice, the creep curve $\gamma(t)$ has the same form as for a network in which breakage is disallowed, recall Fig.~\ref{fig:viscoelastic_sigma}(a), with the  corresponding strain rate  decaying as $\gdot\sim t^{-2}$. At longer times, however, qualitatively different behaviour arises. In particular, in cases where the imposed stress exceeds a critical yield stress $\Sigma_{\rm y}$, the strain  rises slowly at first beyond a threshold value $\gamma_{\rm y}$, then increases dramatically without bound as the network suddenly fails. Significant plastic bond breakage is involved in the failure process. We characterise this by plotting the fraction of broken bonds $f$ in panel d), parametrically as a function of the accumulating strain. We define the threshold strain $\gamma_{\rm th}$ in practice as the limiting strain attained at long times $t\to\infty$ in a simulation performed at the largest stress value for which the sample just avoids failure $\Sigma=\Sigma_{\rm y}^-$, such that the strain indeed approaches this finite constant $\gamma_{\rm th}$ rather than diverging. This is shown by the dotted line in Fig.~\ref{fig:yielding}.

\begin{figure}[!t]
\centering
  \includegraphics[width=\columnwidth]{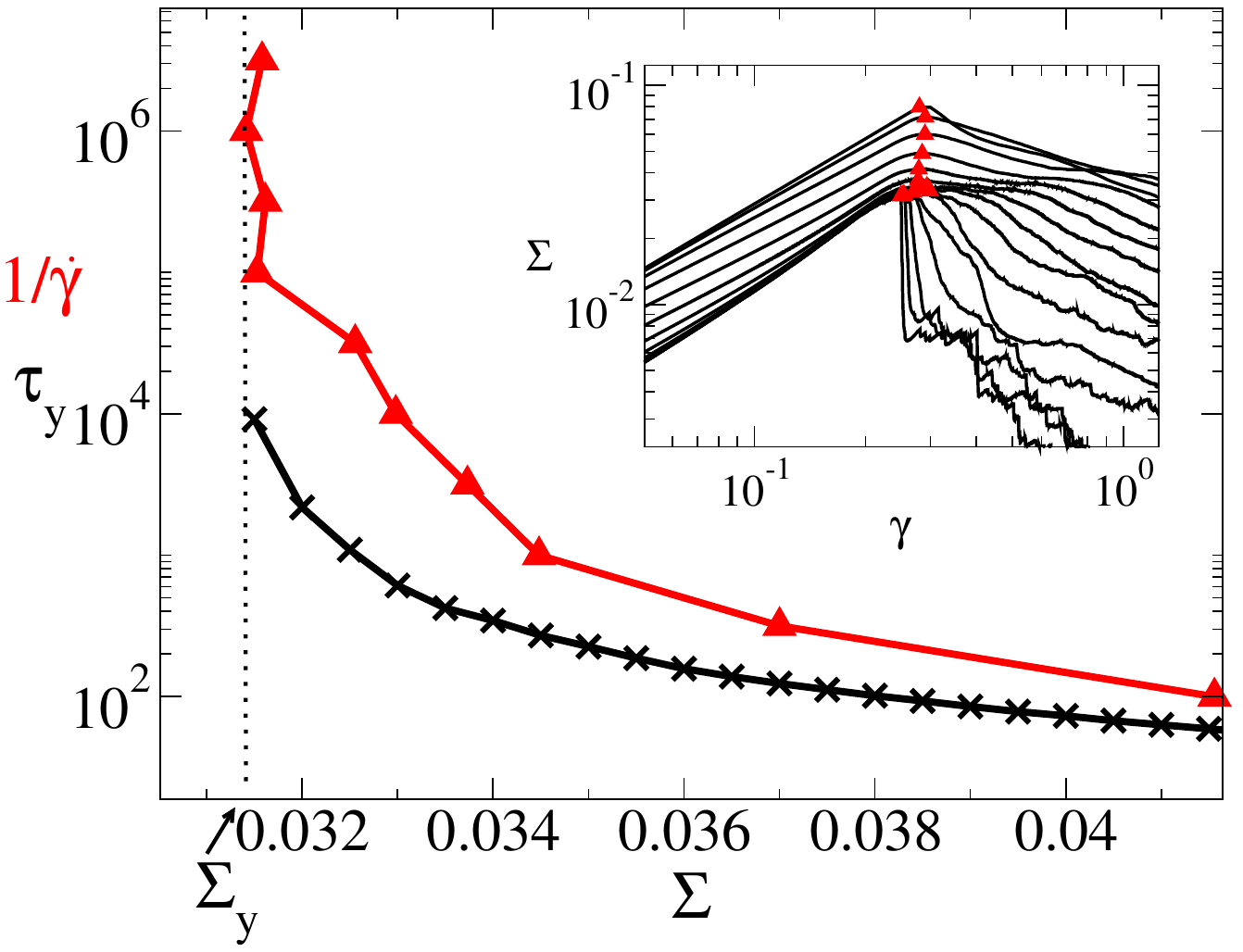}
  \caption{Yielding time $\tau_{\rm y}$ as a function of stress in imposed stress simulations (black crosses), for a filament breakage threshold $b=0.1585$ and initial network connectivity $Z=4.5$. Vertical dotted line shows the yield stress $\Sigma_{\rm y}$, defined as the minimum imposed stress value for which yielding occurs. Inset shows stress versus strain in imposed strain simulations, in which the strain is ramped up from zero at a constant rate $\dot{\gamma}$, for decreasing strain rate values in curves downwards. The stress at which yielding occurs in that imposed strain protocol is identified by the red triangle in each case. These red triangles are then transcribed into the representation of $1/\dot{\gamma}$ versus $\Sigma$ in the main figure. As can be seen, the yield stress observed in the imposed stress simulations agrees well with that obtained in the limit $\dot{\gamma}\to 0$ of the imposed strain simulations.}
  \label{fig:compare}
\end{figure}

For each case in which failure occurs, we define the yielding time $\tau_{\rm y}$ as the time at which the strain first exceeds $1.5$, as indicated by the cross symbols in Fig.~\ref{fig:yielding}. This  is plotted as a function of imposed stress in Fig.~\ref{fig:compare}. As can be seen, $\tau_{\rm y}$ increases dramatically as $\Sigma$ decreases downwards towards a critical value $\Sigma_{\rm y}$. For $\Sigma<\Sigma_{\rm y}$, failure never occurs within any feasible simulation time. For $\Sigma=\Sigma_{\rm y}$, however, $\tau_{\rm y}$ still has  a finite (but large) value. 

A natural question  to ask is to what extent this yield stress, as characterised in our stress-imposed simulations, accords with any yield stress as measured in a strain-imposed deformation protocol. To investigate this, we also performed shear startup simulations, in which a freshly prepared network is subject to the switch-on at time zero of a constant shear rate $\gdot$, with the stress response then measured as a function of accumulating strain $\gamma=\gdot t$. This is reported in the inset to Fig.~\ref{fig:compare}, for several values of imposed strain rate, increasing in curves upward. In each curve, we identify the maximum stress attained before yielding occurs, as indicated by the red triangles. We then plot in the main panel the relationship between this maximum stress and the imposed strain rate in the representation of $1/\gdot$ versus $\Sigma$. In the limit of quasistatic shear $\gdot\to 0$, {\it i.e.} for $1/\gdot\to\infty$ up the vertical axis, the maximum stress attained in shear startup accords with the yield stress measured in our imposed-stress simulations. This gives confidence that the yield stress $\Sigma_{\rm y}$ as measured at imposed stress is the material's true yield stress, and that the yield time $\tau_{\rm y}$ for $\Sigma<\Sigma_{\rm y}$ is truly infinite, rather than simply beyond feasible  run times.

\begin{figure}[!t]
\centering
  \includegraphics[width=1.0\columnwidth]{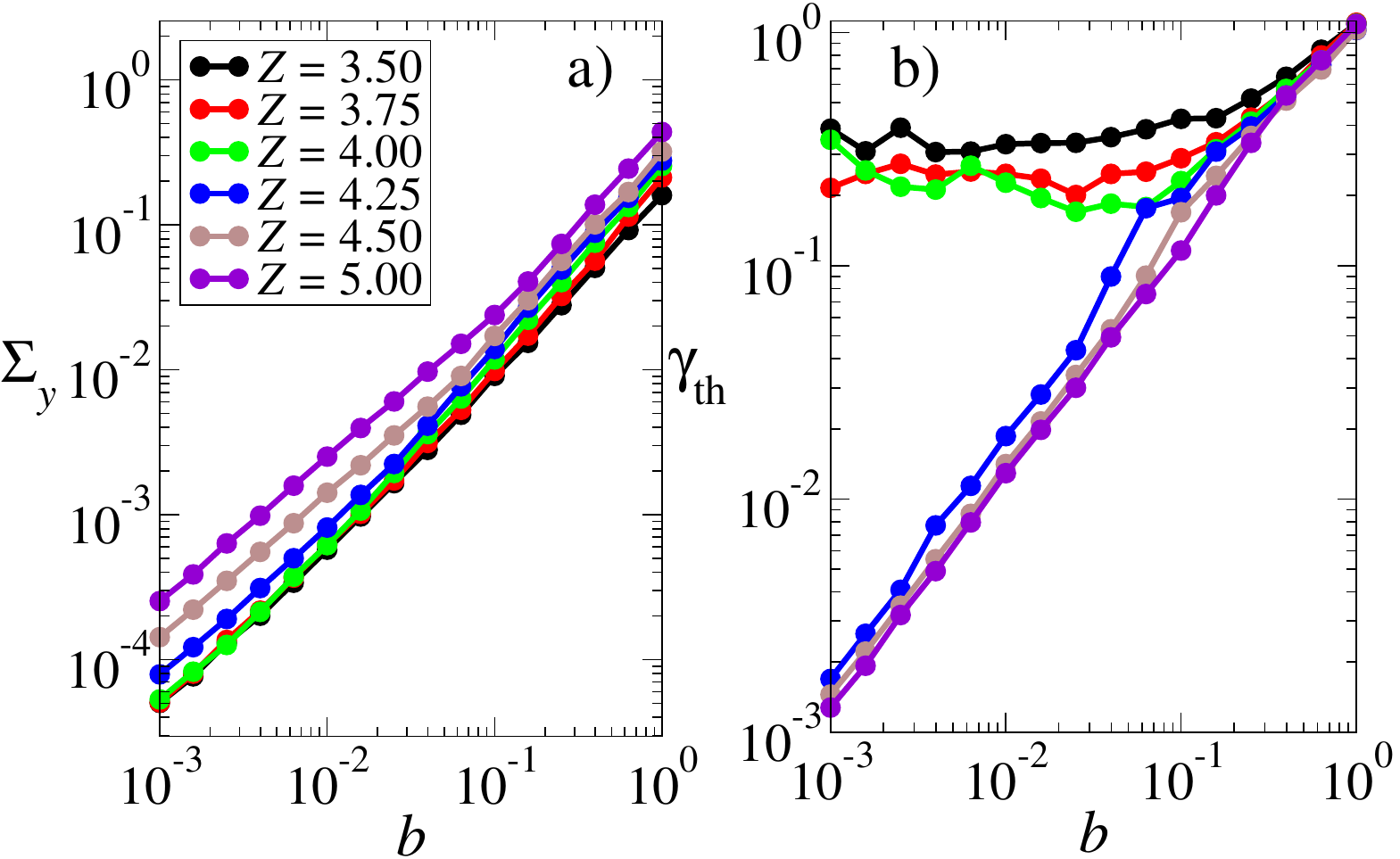}
  \caption{a) Yield stress (as indicated in Fig.~\ref{fig:compare}) and b) threshold strain (as indicated in Fig.~\ref{fig:yielding}) as a function of the filament breakage threshold  $b$ for different values of initial network connectivity $Z$. Both quantities are  defined in the main text. Legend in (a) also applies to (b) with the same colour coding.}
  \label{fig:yieldPoints}
\end{figure}

Having defined the material's threshold strain $\gamma_{\rm th}$ and yield stress $\Sigma_{\rm y}$, we plot these two quantities as a function of the filament breakage threshold in Fig.~\ref{fig:yieldPoints}, for several values of the network connectivity $Z$. As can be seen, the material's yield stress $\Sigma_{\rm y}$ depends in a roughly  linear way on the filament breakage strain $b$ for all values of the connectivity $Z$. In contrast, the threshold strain $\gamma_{\rm th}$ rises linearly with $b$ only for initially stiff networks, $Z>4$. For floppy networks with $Z<4$, in contrast, the threshold strain $\gamma_{\rm th}$ has a finite value even as the filament breakage threshold $b\to 0$, consistent with a vanishingly small stress inducing a strain $O(1)$ in a floppy network.

\subsection{Recoverable and unrecoverable strain}

The degree to which plastic deformation arises while a solid material is under stress has important implications for the extent to which it is later able to recover its original shape if the stress is subsequently removed. This is characterised by the notion of "recoverable strain", defined as the amount of strain recovered by a body after a stress is removed, in the direction opposite to that in which the stress had acted. Strain recovery is generally  understood as being elastic in origin: in a spring network, springs that are elastically stretched in the direction of the imposed stress will recoil after the stress is removed, on a timescale  set by the interplay of their elasticity with dissipative drag. Any unrecoverable strain is instead attributed to the irreversible plastic deformation that arose while the material was under stress. (See however Ref.~\cite{lockwood2025recoverable}.)

\begin{figure}[!t]
\includegraphics[width=1.0\columnwidth]{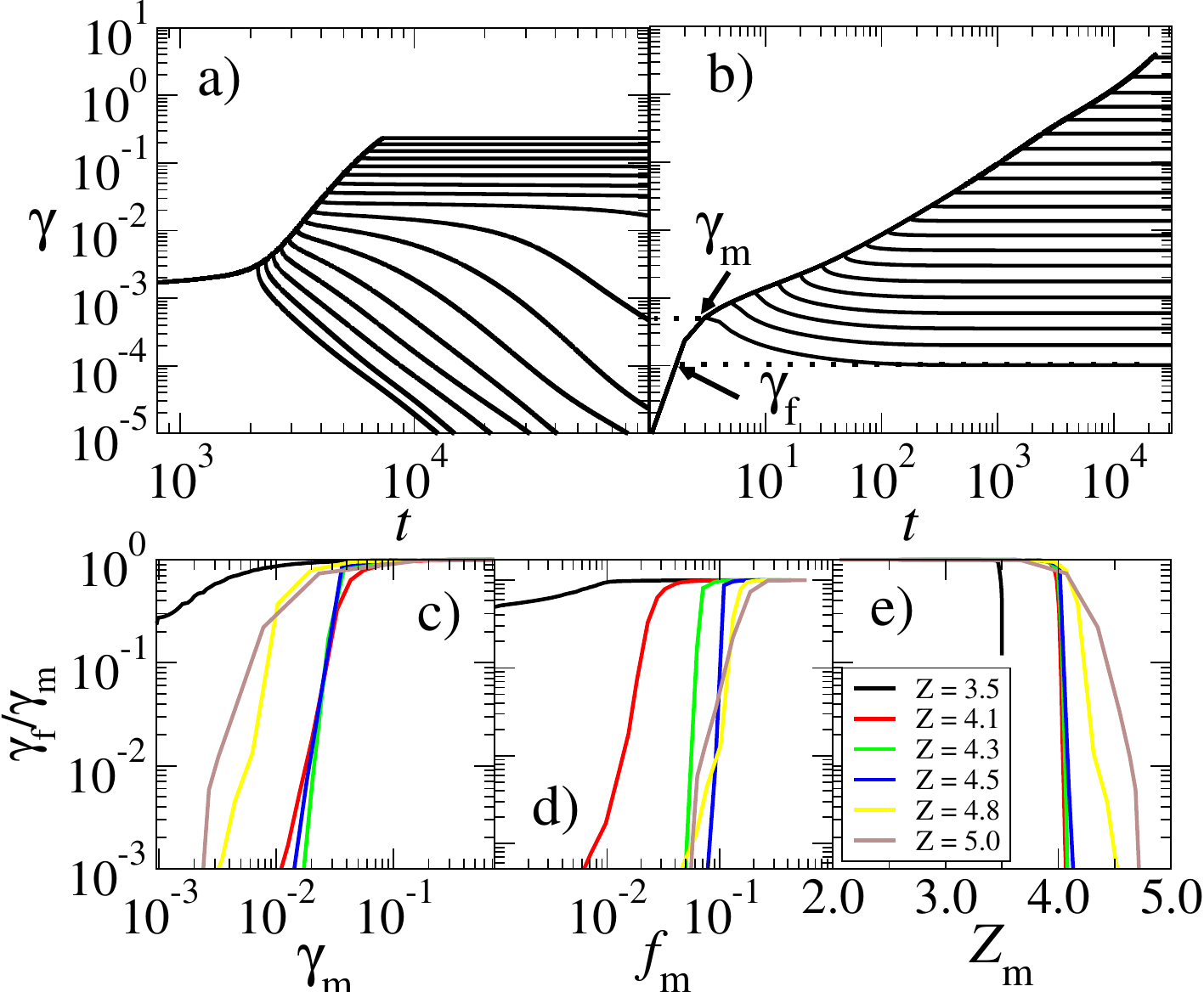}
  \caption{ a)  Strain against time for initial network connectivity $Z = 4.3$, imposed stress $\sigma = 0.0001$ and filament breakage threshold $b = 0.000867$ with the stress imposed for a range of different times before being set back to zero. b) Corresponding data for $Z = 3.5$, $\sigma = 0.0001$ and $b = 0.000837$. Fraction of unrecoverable strain as a function of c) maximum strain attained before unloading and d) fraction of broken filaments at that maximum strain, for several values of (initial) network connectivity increasing in curves downward. e) Fraction of unrecoverable strain as a function of the network connectivity at the maximum strain attained before unloading. Legend in (e) also applies to (c) and (d) with the same colour coding.}
  \label{fig:recovery}
\end{figure}

Fig.~\ref{fig:recovery}(a) shows at early times the increase of strain while a load is imposed on an initially stiff network of connectivity $Z=4.3$. The strain increases up to a maximum value, which exists at the time the load is removed. After the load is removed, the strain decreases as the material  recovers its shape to some extent. Curves upwards correspond to increasing durations of load application prior to removal, all for the same imposed stress. As can be seen, if the stress is removed after only a small amount of creep has occurred, the strain fully reverses to zero after the load is removed and the material exactly recovers its original shape. If the load is instead removed after a much larger amount of forward straining, only part of the strain  subsequently recovers and the material's shape is changed permanently. 

To quantify this, we define $\gamma_{\rm m}$ to be the strain that exists immediately prior to load removal, and $\gamma_{\rm f}$ to be the strain attained at long times after load removal. The quantity $\gamma_{\rm f}/\gamma_{\rm m}$ is then the fraction of strain that is {\em not} recovered after unloading: {\it i.e.} the {\em un}recoverable strain. A value of this fraction equal to zero corresponds to full recovery, whereas a value equal to unity corresponds to no recovery.  Fig.~\ref{fig:recovery}(c) shows this fraction of unrecoverable strain as a function of the strain $\gamma_{\rm m}$ attained immediately prior to load removal, for several values of initial network connectivity. For initially stiff networks with $Z>4$, a non-zero level of unrecoverable strain sets in only beyond some non-zero threshold level of forward straining $\gamma_{\rm m}$. Perhaps more surprising is that it furthermore does so only after a non-zero fraction of springs has broken (panel d): a highly connected network is apparently able to suffer some fraction of its springs failing, yet still fully  recover its shape after load removal. This is seen also in panel e), where the level of unrecoverable strain is plotted as a function of the network connectivity  $Z_{\rm m}$ at the time of load removal, which is lower than its connectivity $Z$ in its freshly prepared state, for any $Z>4$. For an initial connectivity $4.0<Z<4.5$, the strain appears fully recoverable provided the final connectivity of the network at the maximum attained strain under load still exceeds the threshold for rigidity, $Z>4.0$.

In Fig.~\ref{fig:recovery}(b), we plot the forward straining during loading and strain recovery after unloading for a network that was instead prepared in a floppy state,  $Z=3.5$. In this case, even after only small strains have arisen during loading, the strain reverses only partially and material is unable fully to recover its shape. This is to be expected for a floppy network, which lacks the elasticity needed to show a recoil in strain after unloading. Indeed, in such a network, the occurrence of any strain recovery must presumably be attributed to a finite rate of strain developing while the stress is imposed, such that some elasticity builds up in the network despite it still being in the regime  $Z<4$ and $\gamma<\gamma_{\rm c}(Z)$, in which it would be floppy in any quasistatic shearing protocol. This is consistent also with panel d), in which the black curve for $Z=3.5$ shows that the fraction of unrecoverable strain tends to a non-zero value even as the fraction of springs that break during the forward straining tend to zero.

\subsection{Viscoelasticity vs elastoplasticity in power law creep}

Disordered soft solids subject to a constant imposed stress commonly display power law creep, in which the strain rate decreases sublinearly in time, $\gdot\sim t^{n}$  with $-1<n<0$. The strain accordingly increases sublinearly, $\gamma\sim t^{1+n}$.  The particular case $n\to -1$ corresponds to logarithmic creep with $\gamma\sim \log t$ and $\gdot\sim 1/t$. We have already seen one example of power law creep in our network model, in the purely viscoelastic response of a system with coordination at or just above rigidity, $Z\gtrapprox 4.0$, and in which filament breakage is disallowed. Recall Fig.~\ref{fig:power}, which shows power law  creep with an exponent $-1<n_1<0$ over a long duration $t<t_{\rm c}$ for $Z\gtrapprox 4.0$.

Engineering a value of network connectivity in a small window just above rigidity however represents a degree of tuning that seems too specific to represent a phenomenon that is so widely seen experimentally. Indeed, in any case, most gels in practice are much less well connected, typically with $Z<3.0$ in $d=3$ dimensions, for which the marginal value for rigidity is $Z=6.0$. Any counterpart in our $d=2$ simulations would correspond to connectivity values well below $4.0$. For such low connectivities, we did not find power law creep in our purely viscoelastic simulations with filament breakage  disallowed. Recall Fig.~\ref{fig:viscoelastic_sigma}(c,d) and the upper curves in Fig.~\ref{fig:power}(a). These show a  near constant strain rate at early times, while the network is still floppy. After the network has stiffened, the strain rate quickly decreases as $\gdot\sim t^{-2}$ as the strain saturates to its constant value in steady state.

\begin{figure}[!t]
  \includegraphics[width=1.0\columnwidth]{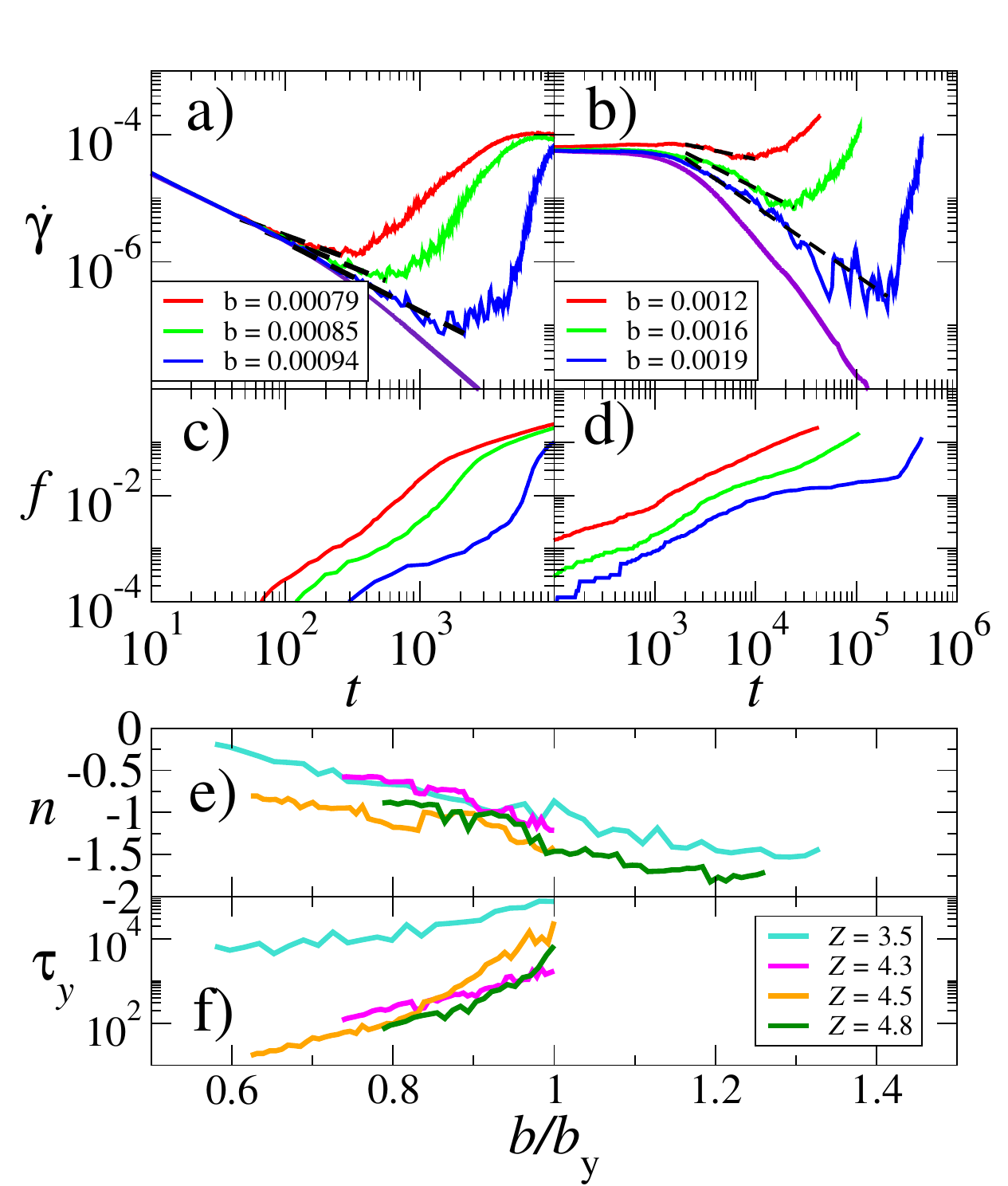}
  \caption{Creep and yielding with filament breakage. a+b) Strain rate versus time for an imposed stress $\Sigma$=10$^{-4}$ and different values of the filament breakage threshold $b$. The initial network connectivity $Z = 4.3$ in a) and $Z = 3.5$ in b). The lowest curve in each of a) and b) corresponds to a network with unbreakable filaments, $b\to \infty$.  To reduce noise in the strain-rate signal, the strain was first coarse grained, being recorded only each time it exceeded the previous recorded value by a multiple $1.005$. This coarse grained strain data was then differentiated to give the curves $\gdot(t)$ shown here. c) and d) show the fraction of broken filaments corresponding to the strain rate curves in a) and b) respectively. The lower panels show as a function of the normalised filament breakage threshold the exponent e) and duration f) of the  power law creep regime $\dot{\gamma}\sim t^{n}$ indicated by black dashed lines in a) and b). To measure the duration of the creep regime, we define its end point by the time at which the strain rate attains its minimum value. The start is defined by the time at which the first filament breaks for $Z>4.0$, and by the constant time $t=2000$ for $Z<4.0$. Legends in (a), (b) and (f) also apply to (c) and (d) and (e) respectively with same colour coding.}
  \label{fig:power1}
\end{figure}

With this in mind, we ask now whether a prolonged regime of power law creep can  emerge in sub-rigid gels with $Z<4.0$, once filament breakage is taken into account. A natural way to explore this would be to perform a series of simulations (for any given connectivity $Z<4.0$) at fixed breakage threshold $b$, for different values of imposed stress $\Sigma$ both below and above the yielding threshold $\Sigma_{\rm y}(b)$ in Fig.~\ref{fig:yieldPoints}(a). This would correspond to taking a vertical cut up that plane of $\Sigma,b$ at fixed $b$. Almost equivalently, we could instead perform a series of simulations at fixed stress $\Sigma$, for different values of breakage threshold both above and below the critical value $b_{\rm y}(\Sigma)$, defined such that networks with $b<b_{\rm y}(\Sigma)$ yield under an imposed stress of $\Sigma$, whereas networks with $b>b_{\rm y}$ do not. This would correspond to taking a horizontal cut across that plane of $\Sigma,b$ at fixed $\Sigma$. Although the former feels more intuitive, given that $\Sigma$ is  more readily varied experimentally than $b$, we actually do the latter. We do so in order to keep the stress  fixed at a small value, which is the regime in which slow creep is most likely to emerge.

\begin{figure}[!t]
  \includegraphics[width=0.5\columnwidth]{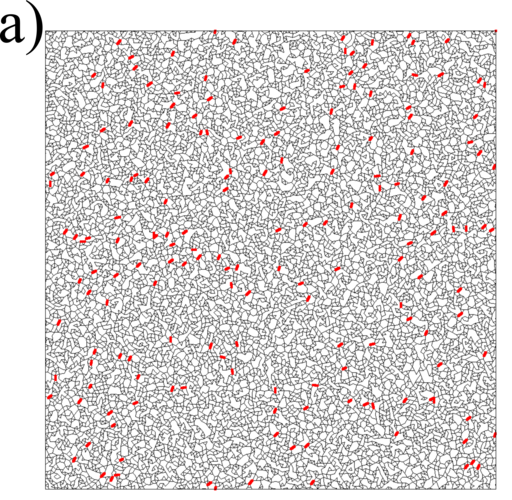}
  \includegraphics[width=0.5\columnwidth]{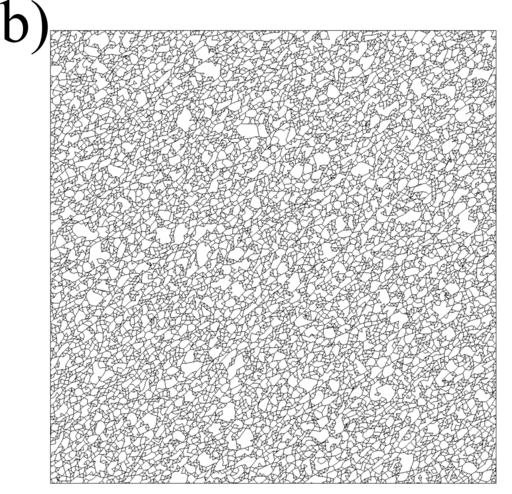}
  \includegraphics[width=0.5\columnwidth]{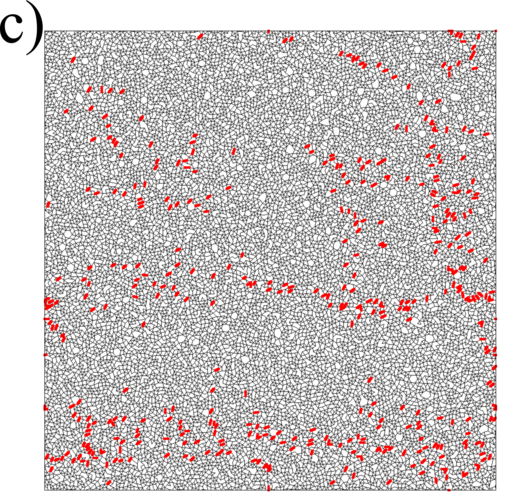}
  \includegraphics[width=0.5\columnwidth]{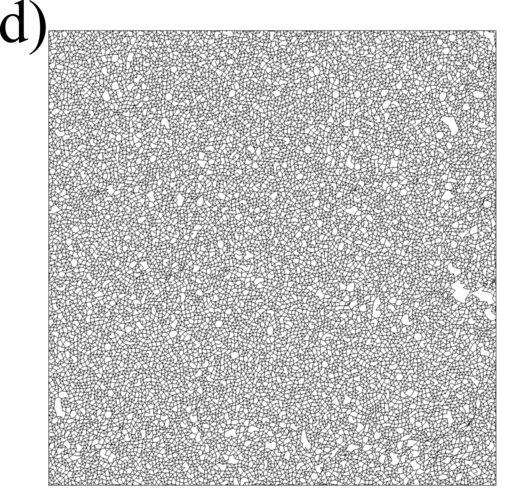}
  \caption{Snapshots of a network under stress. a+b) Initially floppy network with $Z=3.5$. c+d) Initially stiff network with $Z=4.5$. The breakage threshold $b=0.0631$ in each case. a+c) Snapshots in the cosheared frame, with broken bonds coloured red. b+d) Snapshots in the lab frame, with broken bonds removed. In each case, the imposed stress is just above the yield stress, such that each  network will eventually fail completely. Snapshots are shown at the time the fraction of broken bonds first exceeds the fraction of bonds that break in total in a network subject to a stress just below the yield stress.}
  \label{fig:networks}
\end{figure}

Accordingly, the top two panels of Fig.~\ref{fig:power1} show time-differentiated creep curves for several different values of the filament breakage threshold $b$ for a small fixed stress $\Sigma=10^{-4}$. Panel a) pertains to an initially rigid network with $Z=4.3$, and b) to an initially floppy network with $Z=3.5$. Panels c) and d) show the corresponding fraction of broken springs as a function of time. For each time-differentiated creep curve shown in a) and b), we perform a best fit to the power law regime $\gdot\sim t^{n}$ to extract values of the creep exponent $n$, along with the duration $\tau_{\rm y}$ of the creep regime. These quantities are plotted as a function of the normalised filament breakage threshold $b/b_{\rm y}$ in panels e) and f), for several different values of $Z$.

Particularly interesting are the results for the subcritical network with $Z=3.5$, in the regime where the network eventually fails, $b/b_{\rm y}<1.0$. In this regime, the duration of the creep regime is large, $\tau_{\rm y}=10^4-10^5$, and the creep exponent  $-1<n<0$.  A sub-rigid gel with filament breakage accordingly indeed shows a prolonged regime of power law creep in which the strain rate decreases sublinearly over time.  Interestingly, the threshold case $b=b_{\rm y}$ appears to correspond to a creep exponent $n=-1$, to within the noise of our data. Here we have explored only a single value of imposed stress $\Sigma=10^{-4}$, for which the creep regime  has an extremely long duration. It would be interesting in future work to explore whether the duration $\tau_{\rm y}$ of the creep regime increases even further with decreasing imposed stress. Such simulations are however extremely time consuming computationally.

Finally, we show snapshots of a network under stress in Fig.~\ref{fig:networks}. The top two panels represent an initially floppy network with $Z=3.5$. The bottom two represent an initially stiff network with $Z=4.5$. In each case, the left panel shows a snapshot in the cosheared frame, with broken springs indicated in red. The right panel shows a snapshot in the lab frame, with broken springs removed. As can be seen, the bond breakages in a stiff network tend to localise in macroscopic bands that span the network. In contrast, in a floppy network it is spread much more diffusively across the system. It would be interesting in future work to explore further this strain localisation during the yielding of initially rigid networks.

\section{Conclusions}
\label{sec:conclusions}

In this work, we have developed a numerical method to simulate a minimal athermal fibre network model subject to a constant imposed shear stress, and used it to study the creep and yielding of disordered gels and fibre network materials. 

Networks in which filament breaking is disallowed show purely viscoelastic creep, without any elastoplastic deformation. In this case, a stiff network with coordination $Z>2d$ exhibits time-differentiated creep curves  that show scaling collapse when plotted as $\gdot(t)/\Sigma$ vs $t$, for several small values of imposed stress. For a floppy network with coordination $Z<2d$, the differentiated creep curves instead show scaling collapse when plotted as $\gdot(t)/\Sigma$ vs $\Sigma t$. The corresponding creep curve $\gamma(t)$ then depends on stress only via the rescaled time variable $\Sigma t$, rather than having an amplitude proportional to the stress, marking a significant departure from the behaviour of a Hookean solid. In both stiff and floppy networks, viscoelastic creep terminates as the strain approaches a limiting steady state value, with the strain rate decreasing as $\gdot\sim t^{-2}$. The steady state strain is linear in the imposed stress for stiff networks, but shows a non-zero intercept in the limit of small stress for floppy networks. For marginally connected networks with $Z\gtrapprox 4.0$, a prolonged regime of viscoelastic power law creep is seen before the strain saturates, with the strain increasing sublinearly and the strain rate decaying sublinearly as a function of time.

For a network in which filament breakage is allowed, with a finite breakage threshold $b$, there exists a critical imposed stress  $\Sigma_{\rm y}(b)$ above which network will always eventually catastrophically fail. The time taken to approach failure however increases dramatically with decreasing imposed stress $\Sigma\to\Sigma_{\rm y}^+$. In this regime, an initially floppy network shows a prolonged regime of power law elastoplastic creep in which filaments progressively  break, with the strain rate decreasing sublinearly over time. We  have further explored the implications of viscoelastic and elastoplastic creep for strain recovery after a material is unloaded, following an interval of creep under an applied load.

Importantly, we have reported two distinct regimes of  power law creep in which the strain increases sublinearly and the strain rate decreases sublinearly over time, with a distinct physical mechanism in each regime.  The first regime arises in networks with marginal connectivity, $Z\gtrapprox 4.0$, is purely viscoelastic in nature, and arises from non-affine internal deformations within the network, but without filament breakage. The second arises also in initially floppy networks with $Z<4.0$ and is elastoplastic in nature, entailing filament breakage.   Whether these two mechanisms are together sufficient to explain why power law creep appears as so widespread a phenomenon across network materials remains an open question. In future work, it would be interesting to explore additional possible explanations such as  Rouse-like motion of network chains~\cite{rubinstein2003polymer}, a fractal underlying network structure~\cite{head2022viscoelastic}, and thermal effects such as network linker dynamics~\cite{broedersz2010cross}. 


\section*{Conflicts of interest}
There are no conflicts to declare.

\section*{Acknowledgements}
This project has received funding from the European Research Council (ERC) under
the European Union’s Horizon 2020 research and innovation programme (grant agreement No. 885146).

\balance


\bibliographystyle{rsc} 

\providecommand*{\mcitethebibliography}{\thebibliography}
\csname @ifundefined\endcsname{endmcitethebibliography}
{\let\endmcitethebibliography\endthebibliography}{}

\end{document}